\documentclass[12pt]{article}
\usepackage{mathptmx}
\usepackage{cite} 
\usepackage[scaled=0.9]{helvet}
\usepackage{graphicx}
\usepackage{amsmath}

\usepackage{url}  
\usepackage{array}

\evensidemargin\oddsidemargin
\parskip=\medskipamount

\makeatletter
\long\def\@makecaption#1#2{%
  \vskip\abovecaptionskip
  \sbox\@tempboxa{\small{\bfseries #1} \  #2}%
  \ifdim \wd\@tempboxa >\hsize
    \small{\bfseries #1} \  #2\par
  \else
    \global \@minipagefalse
    \hb@xt@\hsize{\hfil\box\@tempboxa\hfil}%
  \fi
  \vskip\belowcaptionskip}
\renewcommand\section{\@startsection {section}{1}{\z@}%
      {-3.25ex\@plus -1ex \@minus -.2ex}%
      {1ex \@plus .2ex}%
      {\normalfont\large\sffamily\bfseries}}
\renewcommand\subsection{\@startsection{subsection}{2}{\z@}%
      {-3ex\@plus -1ex \@minus -.2ex}%
      {0.5ex \@plus .2ex}%
      {\normalfont\normalsize\sffamily\bfseries}}
\renewcommand\subsubsection{\@startsection{subsubsection}{3}{\z@}%
      {-3ex\@plus -1ex \@minus -.2ex}%
      {0.25ex \@plus .2ex}%
      {\normalfont\normalsize\sffamily\bfseries}}
\renewcommand\paragraph{\@startsection{paragraph}{4}{\z@}%
      {3ex \@plus1ex \@minus.2ex}%
      {-1em}%
      {\normalfont\normalsize\sffamily\bfseries}}
\renewcommand\subparagraph{\@startsection{subparagraph}{5}{\z@}%
      {1ex \@plus.5ex \@minus .2ex}%
      {-1em}%
      {\normalfont\normalsize\sffamily\bfseries}}
\makeatother

\def\qsqmax{q^2_\mathrm{max}}
\def\sth{s_\mathrm{th}}
\def\mbstar{m_{B^*}}
\def\modvub{|V_{ub}|}
\def\n#1e#2n{#1\times10^{#2}}
\def\gev{\,\mathrm{GeV}}

\def\vubresult{(3.47\pm0.29) \times 10^{-3}}

\begin{document}

\begin{center}\Large\bfseries\sffamily
$\modvub$ from Exclusive Semileptonic $B\to\pi$ Decays Revisited
\end{center}

\begin{center}
\textbf{\textsf{Jonathan M Flynn${}^\mathrm{a}$ and Juan
  Nieves${}^\mathrm{b}$}}\\[2ex]
${}^\mathrm{a}$School of Physics and Astronomy, University of
  Southampton\\
  Highfield, Southampton SO17~1BJ, UK\\
${}^\mathrm{b}$Departamento de F\'isica At\'omica, Molecular y
  Nuclear, Universidad de Granada,\\
  E--18071 Granada, Spain
\end{center}
\medskip

\begin{quote}
\begin{center}\textbf{\textsf{Abstract}}\end{center}
We update the extraction of $\modvub$ from exclusive semileptonic
$B\to\pi$ decays, combining experimental partial branching fraction
information with theoretical form factor calculations, using the
recently revised HPQCD results for the form factors $f_+$ and $f_0$.
We use Omn\`es representations to provide the required
parametrisations of the form factors. The extracted value is
$\modvub=(3.47\pm0.29\pm0.03)\times10^{-3}$, in striking agreement
with $\modvub$ extracted using all other inputs in CKM fits and
showing some disagreement with $\modvub$ extracted from inclusive
semileptonic $B\to\pi$ decays.
\end{quote}

In this short note we update our extraction of $\modvub$ from combined
experimental and theoretical information on exclusive semileptonic
$B\to\pi$ decays in light of the recently revised values for the form
factors $f_+$ and $f_0$ from the lattice QCD calculation by the HPQCD
collaboration~\cite{Dalgic:2006dt}. Our analysis procedure and
inputs are fully described in~\cite{Flynn:2006vr,Flynn:2007qd}. We
combine experimental partial branching fraction information with
theoretical calculations of both form factors, using Omn\`es
representations to provide parametrisations of the form factors. The
Omn\`es representation for $f_+(q^2)$ takes into account the existence
of the $B^*$ pole as described in~\cite{Flynn:2007qd}.

We have used experimental partial branching fraction data from the
tagged analyses of CLEO~\cite{Athar:2003yg},
Belle~\cite{Hokuue:2006nr} and BaBar~\cite{Aubert:2006ry}, and from
the untagged analysis of BaBar~\cite{Aubert:2006px,BaBar:EPAPS}. When
computing partial branching fractions, we have used $\tau_{B^0}=
1/\Gamma_\mathrm{Tot} = \n(1.527\pm
0.008)e-12n\,\mathrm{s}$~\cite{Barberio:2007cr} for the $B^0$
lifetime. For theoretical form-factor inputs we use the lightcone
sumrule (LCSR) result $f_+(0)=f_0(0) =
0.258\pm0.031$~\cite{LCSR_04_BZ} and lattice QCD results from
FNAL-MILC~\cite{Okamoto:2004xg,Okamoto:2005zg,Mackenzie:2005wu,VandeWater:2006aa}
(using the three $f_+(q^2)$ values quoted in~\cite{Arnesen:2005ez} and
reading off three values for $f_0(q^2)$ at the same $q^2$ points
from~\cite{Okamoto:2005zg}). The lattice QCD results from HPQCD have
recently been revised~\cite{Dalgic:2006dt} and we note the updated
HPQCD form factor values in table~\ref{tab:HPQCDmod}\footnote{The
changes to the results for $f_0$ are relatively small so we do not
expect large effects on analyses based on these values alone, for
example, using $f_0$ input to extract phase-shift information for
$s$-wave elastic $B\pi$ scattering~\cite{Flynn:2007ki}.}.

Our fit uses four evenly-spaced Omn\`es subtraction points for each
form factor, covering the range $0 \leq q^2 \leq
\qsqmax=(m_B-m_\pi)^2$, together with the value of $\modvub$. The
best-fit parameters are:
\begin{equation}
\label{eq:best-fit}
\begin{array}{rcl}
\modvub         &=& \vubresult \\
f_+(0)=f_0(0)   &=& 0.245\pm0.023 \\
f_+(\qsqmax/3)  &=& 0.475\pm0.046 \\
f_+(2\qsqmax/3) &=& 1.07 \pm0.08 \\ 
f_+(\qsqmax)    &=& 7.73 \pm1.29 \\
f_0(\qsqmax/3)  &=& 0.338\pm0.089 \\
f_0(2\qsqmax/3) &=& 0.520\pm0.041 \\ 
f_0(\qsqmax)    &=& 1.06 \pm0.26
\end{array}
\end{equation}
The fit has $\chi^2/\mathrm{dof} = 0.74$ for $24$ degrees of freedom,
while the Gaussian correlation matrix of fitted parameters is:
\begin{equation}
\left(
\begin{array}{cccccccc}
1 & -0.39 & -0.92 & -0.83 & -0.56 & -0.02 &  0.00 &  0.00 \\
  &  1    &  0.19 &  0.48 & -0.04 &  0.06 &  0.00 & -0.01 \\
  &       &  1    &  0.77 &  0.59 &  0.01 &  0.00 &  0.00 \\
  &       &       &  1    &  0.38 &  0.03 &  0.00 &  0.00 \\
  &       &       &       &  1    &  0.00 &  0.00 &  0.00 \\
  &       &       &       &       &  1    &  0.25 &  0.83 \\
  &       &       &       &       &       &  1    &  0.21 \\
  &       &       &       &       &       &       &  1
\end{array}
\right)
\end{equation}

\begin{figure}
\begin{center}
\includegraphics[width=\hsize]{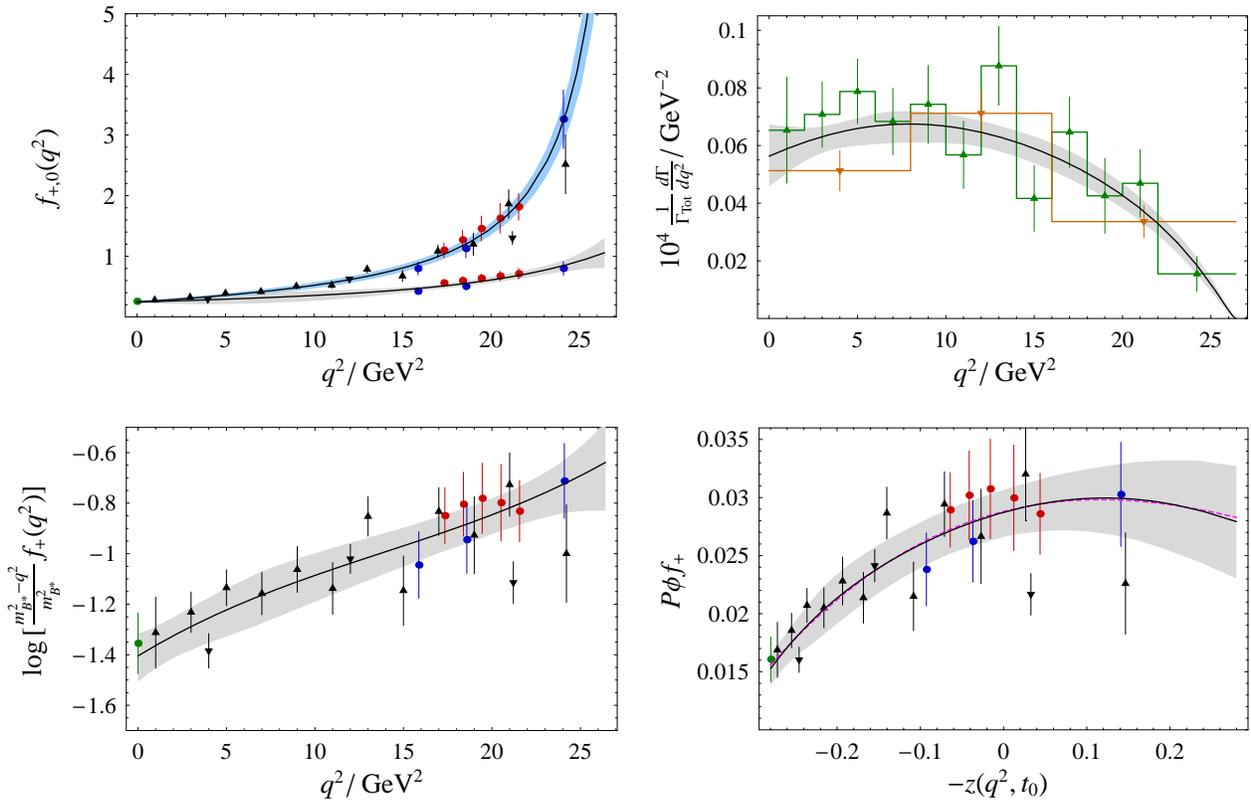}
\end{center}
\caption{Results obtained from the fit to experimental partial
  branching fraction data and theoretical form factor calculations.
  The top left plot shows the two form factors with their error bands,
  the lattice and LCSR input points (dots: green LCSR, red HPQCD, blue
  FNAL-MILC) and `experimental' points (black triangles,
  upward-pointing for tagged and downward pointing for untagged data)
  constructed by plotting at the centre of each bin the constant form
  factor that would reproduce the partial branching fraction in that
  bin. The top right plot shows the differential decay rate together
  with the experimental inputs. The bottom plots provide more details
  of the inputs and fits by showing on the left
  $\log[(\mbstar^2-q^2)f_+(q^2)/\mbstar^2]$ as a function of $q^2$,
  and on the right $P\phi f_+$ as a function of $-z$. The dashed
  magenta curve in the bottom right plot is a cubic polynomial fit in
  $z$ to the Omn\`es curve.}
\label{fig:results}
\end{figure}

In figure~\ref{fig:results} we show the fitted form factors, the
differential decay rate calculated from our fit and the quantities
$\log[(\mbstar^2-q^2)f_+(q^2)/\mbstar^2]$ and $P\phi
f_+$~\cite{Hill:2006ub} where the details of the fit and inputs can
better be seen. The dashed magenta curve in the $P\phi f_+$ plot is a
cubic polynomial fit in $z$ to the output from our
analysis\footnote{Expressions for $P$, $\phi$ and $z$ can be found
in~\cite{Arnesen:2005ez}. We set $t_0 =
\sth(1-\sqrt{1-\qsqmax/\sth})$, where $\sth=(m_B+m_\pi)^2$, which is
the `preferred choice', labelled BGLa, in~\cite{Ball:2006jz}. This
choice for $t_0$ ensures that $|z|\leq0.3$ for $0\leq
q^2\leq\qsqmax$.}. The sum of squares of the coefficients in this
polynomial is $A = \sum_n a_n^2=0.012$ which is consistent with being
of order $(\Lambda/m_b)^3$~\cite{Becher:2005bg}, where $\Lambda$ is a
hadronic scale and $m_b$ is the $b$-quark mass, and safely satisfies
the dispersive constraint $ A \leq 1$~\cite{Arnesen:2005ez}.
\begin{table}
\begin{center}
\begin{tabular}{>{$}c<{$}>{$}c<{$}>{$}c<{$}}
\hline
\vrule height2.5ex depth0pt width0pt
q^2/\gev^2 & f_+(q^2) & f_0(q^2) \\
\hline
17.34 & 1.101\pm0.053 & 0.561\pm0.026 \\
18.39 & 1.273\pm0.099 & 0.600\pm0.021 \\
19.45 & 1.458\pm0.142 & 0.639\pm0.023 \\
20.51 & 1.627\pm0.185 & 0.676\pm0.041 \\
21.56 & 1.816\pm0.126 & 0.714\pm0.056 \\
\hline
\end{tabular}
\end{center}
\caption{Revised HPQCD results for the form factors $f_+$ and
  $f_0$~\cite{Dalgic:2006dt}. The error shown is statistical only: the
  systematic error for each input form factor value is $10\%$.}
\label{tab:HPQCDmod}
\end{table}

From our fit we calculate the total branching fraction
\begin{equation}
\mathrm{B}(B^0\to\pi^- l^+\nu)=
 (1.37\pm0.08\pm0.01)\times10^{-4}
\end{equation}
where the first uncertainty is from our fit and the second is from the
uncertainty in the experimental $B^0$ lifetime. We evaluate
\begin{equation}
\frac1{m_B} \left.\frac{f_+(\qsqmax)}{f_0(\qsqmax)}\right|_{B\pi}
  = 1.4\pm0.4\gev^{-1}
\end{equation}
to be compared to the corresponding quantity in $D\to\pi$ exclusive
semileptonic decays, $1.4\pm0.1\gev^{-1}$ extracted from the
unquenched lattice QCD results in~\cite{Aubin:2004ej}. We also
calculate the combination,
\begin{equation}
\modvub f^+(0) = (8.5\pm0.8)\times10^{-4}.
\end{equation}
A model-independent extraction of this combination can be performed by
applying soft collinear effective theory (SCET) to $B\to\pi\pi$ decays
and deriving a factorisation result~\cite{vubfplus-fact-2004}. Our
result compares well with $\modvub f^+(0) = \n(7.6\pm1.9)e-4n$ quoted
in~\cite{Stewart:ckm2006} using the SCET/factorisation approach.

We have assumed that the lattice input form factor data have
independent statistical errors and fully-correlated systematic errors
(but no correlations linking $f_+$ and $f_0$). Since we do not know
these correlations we have also performed fits with no correlations in
the lattice inputs and assuming correlated systematic errors linking
$f_+$ and $f_0$. We find that the central fitted value for $\modvub$
shifts by less than $0.03\times10^{-3}$, which we will apply as a
systematic error for our extracted value:
\begin{equation}
\label{eq:vubresult}
\modvub=(3.47\pm0.29\pm0.03)\times10^{-3}.
\end{equation}
This value differs by more than one standard deviation from the
$\modvub$ values extracted from inclusive semileptonic $B\to\pi$
decays and quoted in~\cite{Barberio:2007cr}. However, using the
inclusive determinations with the highest efficiency and best
theoretical control leads to $\modvub =
(4.10\pm0.30_\mathrm{exp}\pm0.29_\mathrm{th})
\times10^{-3}$~\cite{Neubert:FPCP2007} which is consistent with the
value found here.

The result is in very good agreement with values for $\modvub$ coming
from CKM fits using inputs apart from $\modvub$ itself. For example
the angles-only fit in~\cite{Bona:2006ah} leads to
$\modvub=(3.67\pm0.24)\times 10^{-3}$, while the UTfit collaboration's
result for $\modvub$ determined from all other inputs, including
Winter 2007 updated information~\cite{UTfit:web} is
$\modvub=(3.44\pm0.16)\times 10^{-3}$.

The revised HPQCD results are in closer agreement with the FNAL-MILC
results and lead to smaller $\modvub$. These groups use different
methods for treating heavy quarks in their simulations, so the
agreement is very encouraging. However, since they both use the same
input gauge field ensembles, it remains very important that the
outputs are confirmed by independent simulations. Both lattice QCD and
light cone sum-rules calculations of the $B\to\pi$ form factors, when
combined with experimental partial branching fraction information, now
agree on values of $\modvub$ around $3.5 \times 10^{-3}$ or so (see
equation~\eqref{eq:vubresult} and also~\cite{Arnesen:2005ez}
and~\cite{Ball:2006jz}), in striking agreement with the value obtained
using all other inputs in global CKM fits. The hints of a disagreement
with inclusive determinations of $\modvub$ are strengthened.

\subsubsection*{Acknowledgements}

We thank Junko Shigemitsu for communicating the revised HPQCD results
to us and Richard Hill for a comment on the $P\phi f_+$ polynomial
fit. JMF acknowledges the hospitality of the Departamento de F\'isica
At\'omica, Molecular y Nuclear, Universidad de Granada, MEC Grant
SAB2005-0163, and PPARC grant PP/D000211/1. JN acknowledges support
from Junta de Andalucia grant FQM0225 and MEC grant FIS2005--00810.
JMF and JN acknowledge support from the EU Human Resources and
Mobility Activity, FLAVIAnet, contract number MRTN--CT--2006--035482.

\bibliographystyle{physrev}
\bibliography{omnes2}

\end{document}